\documentclass[aps,prl,twocolumn,showpacs,floatfix,superscriptaddress]{revtex4}
\usepackage{graphicx}
\begin{document}

\title{Superconducting density of states and vortex cores of 2H-NbS$_{2}$}

\author{I. Guillam\'on}
\affiliation{Laboratorio de Bajas Temperaturas, Departamento de
F\'isica de la Materia Condensada, Instituto de Ciencia de
Materiales Nicol\'as Cabrera, Facultad de Ciencias \\ Universidad
Aut\'onoma de Madrid, E-28049 Madrid, Spain}
\author{H. Suderow}
\affiliation{Laboratorio de Bajas Temperaturas, Departamento de
F\'isica de la Materia Condensada, Instituto de Ciencia de
Materiales Nicol\'as Cabrera, Facultad de Ciencias \\ Universidad
Aut\'onoma de Madrid, E-28049 Madrid, Spain}
\author{S. Vieira}
\affiliation{Laboratorio de Bajas Temperaturas, Departamento de
F\'isica de la Materia Condensada, Instituto de Ciencia de
Materiales Nicol\'as Cabrera, Facultad de Ciencias \\ Universidad
Aut\'onoma de Madrid, E-28049 Madrid, Spain}
\author{L. Cario}
\affiliation{Institut des Mat\'eriaux Jean Rouxel (IMN), Universit\'e de Nantes - CNRS, 2 rue de la Houssini\'ere, BP 32229, 44322 Nantes Cedex 03, France}
\author{P. Diener}
\affiliation{Institut des Mat\'eriaux Jean Rouxel (IMN), Universit\'e de Nantes - CNRS, 2 rue de la Houssini\'ere, BP 32229, 44322 Nantes Cedex 03, France}
\author{P. Rodi\`ere}
\affiliation{Institut N\'{e}el, CNRS / UJF, 25, Av. des Martyrs,
BP166, 38042 Grenoble Cedex 9, France}

\begin{abstract}

Scanning tunneling microscopy and spectroscopy (STM/S) measurements in the superconducting dichalcogenide 2H-NbS$_{2}$ show a peculiar superconducting density of states with two well defined features at 0.97 meV and 0.53 meV, located respectively above and below the value for the superconducting gap expected from single band s-wave BCS model ($\Delta$=1.76k$_BT_c$=0.9 meV). Both features have a continuous temperature evolution and disappear at $T_{c}$ = 5.7 K. Moreover, we observe the hexagonal vortex lattice with radially symmetric vortices and a well developed localized state at the vortex cores. The sixfold star shape characteristic of the vortex lattice of the compound 2H-NbSe$_{2}$ is, together with the charge density wave order (CDW), absent in 2H-NbS$_{2}$.
\end{abstract}

 \date{\today}

\maketitle

The study of the coexistence of superconductivity with competing physical phenomena such as magnetic or charge order has historically produced great interest on the scientific community. Anisotropies or modulations of the superconducting properties (in real and/or reciprocal space) often appear as a consequence of competing orders within the same system\ \cite{Tsuei00,Flouquet02}. In the compound 2H-NbSe$_{2}$, superconductivity appears within a CDW state ($T_{CDW}$=33K and $T_c$=7.2 K)\ \cite{Moncton77}. Low lying excitations measured deep in the superconducting state long time ago by specific heat\ \cite{Garoche76,Kobayashi77} have been explained by recent experiments and theoretical calculations with a multiband superconductivity and a peculiar anisotropy of the superconducting gap\ \cite{Fletcher07,Rodrigo04c}. Recent angular resolved photoemission spectroscopy measurements demonstrate that the superconducting gap has, close to T$_c$ (at 5.7 K), largest values at k-space positions connected with CDW wavevectors\ \cite{Yokoya01,Kiss07}. Hess et al.\ \cite{Hess89,Hess90,Hess91} found that the local superconducting density of states (LDOS) at the center of the vortex core shows a high peak close to the Fermi level highlighting the lowest quasiparticle state bound within the vortex core well\ \cite{Caroli64}. Around the vortex core, the LDOS is far from respecting in-plane symmetry and intriguing vortex lattice images with patterns showing strong in-plane LDOS modulations are obtained \ \cite{Hess89,Hess90,Hess91}.

2H-NbSe$_2$ belongs to the transition-metal dichalcogenides (2H-MX$_{2}$ with M = Ta, Nb and X = Se, S), a family of systems which is unique to study the interplay between CDW order and superconductivity. The 2H-MX$_{2}$ compounds share a double layered structure made of two hexagonal X sheets with an intercalated M sheet (X-M-X), connected through very weak van der Waals bonds\ \cite{Wilson75}. This produces highly anisotropic, quasi two dimensional electronic properties. The features of the Fermi Surface (FS) expected to be common in all systems of the series are two concentric cylindrical FS sheets centered on both $\Gamma$ and $K$ points, derived from the transition-metal d bands\ \cite{Corcoran94,Suzuki05,Johannes06,Inosov08}. When going over the series from 2H-TaSe$_{2}$ and 2H-TaS$_{2}$ to 2H-NbSe$_{2}$ and 2H-NbS$_{2}$, the ratio of the intralayer lattice constant with the interlayer distance a/c increases, as well as $T_c$, whereas $T_{CDW}$ decreases strongly\ \cite{Wilson75,CastroNeto01,Suderow05d}. In 2H-NbS$_2$, CDW order seems to be absent from macroscopic measurements and $T_c$ is around 6 K \ \cite{Wilson75,Hamaue86}, very close to $T_c$ of 2H-NbSe$_2$ (7.2 K). The study and characterization of superconducting properties of 2H-NbS$_{2}$ arises as an extraordinary route to give new insight into the unsolved issues raised by the peculiar behavior of 2H-NbSe$_2$. Among these issues we highlight the following: Is two band superconductivity common to both compounds? How do quasiparticles behave in the vortex core wells of both compounds? And finally (possibly the most intriguing question), what is, if any, the relationship between the star shaped vortex cores in 2H-NbSe$_2$ and CDW order? Here we discuss STM/S measurements of the superconducting properties of 2H-NbS$_{2}$ which give relevant answers to these questions.

We use a homebuild STM/S system installed in a dilution refrigerator. The experimental setup, as described in previous work\ \cite{Suderow04,Crespo06a,Guillamon07,Guillamon08},
is mechanically decoupled of external vibrations, has a resolution in energy of 15 $\mu$eV that has been previously tested by measuring clean spectra in low $T_{c}$ superconductors as such Al, and a mechanical pulling mechanism that allows to move the tip with respect to the sample holder. 2H-NbS$_2$ was synthesized as described elsewhere\ \cite{Fisher80}. Thousands of layered crystals with different shapes and sizes, typically of some tens of $\mu$m$^2$ were obtained. The chemical composition of these crystals was checked using a scanning electron microscope JEOL 5800 equipped with a microanalyzer. From powder X Ray diffraction technique, the 2H structure was found to be the dominant polytypic form of the batch and only a very small amount of 3R-NbS$_2$  was detectable. Single crystal diffraction experiments performed at room temperature using a four circles FR 590 Nonius CAD-4F Kappa-CCD diffractometer confirmed the presence of the 2H polytype and revealed that the platelets with a well defined hexagonal shape have less stacking faults along the c-axis than the others. For the present experiment, we have chosen three small single crystals with a nice hexagonal shape, which were individually mounted on a gold sample. It was possible to move easily the scanning window from sample to sample crossing a clean surface of gold. The superconducting features discussed below are well reproducible in all three samples.

\begin{figure}[ht]
\includegraphics[height=9cm,clip,angle=-90]{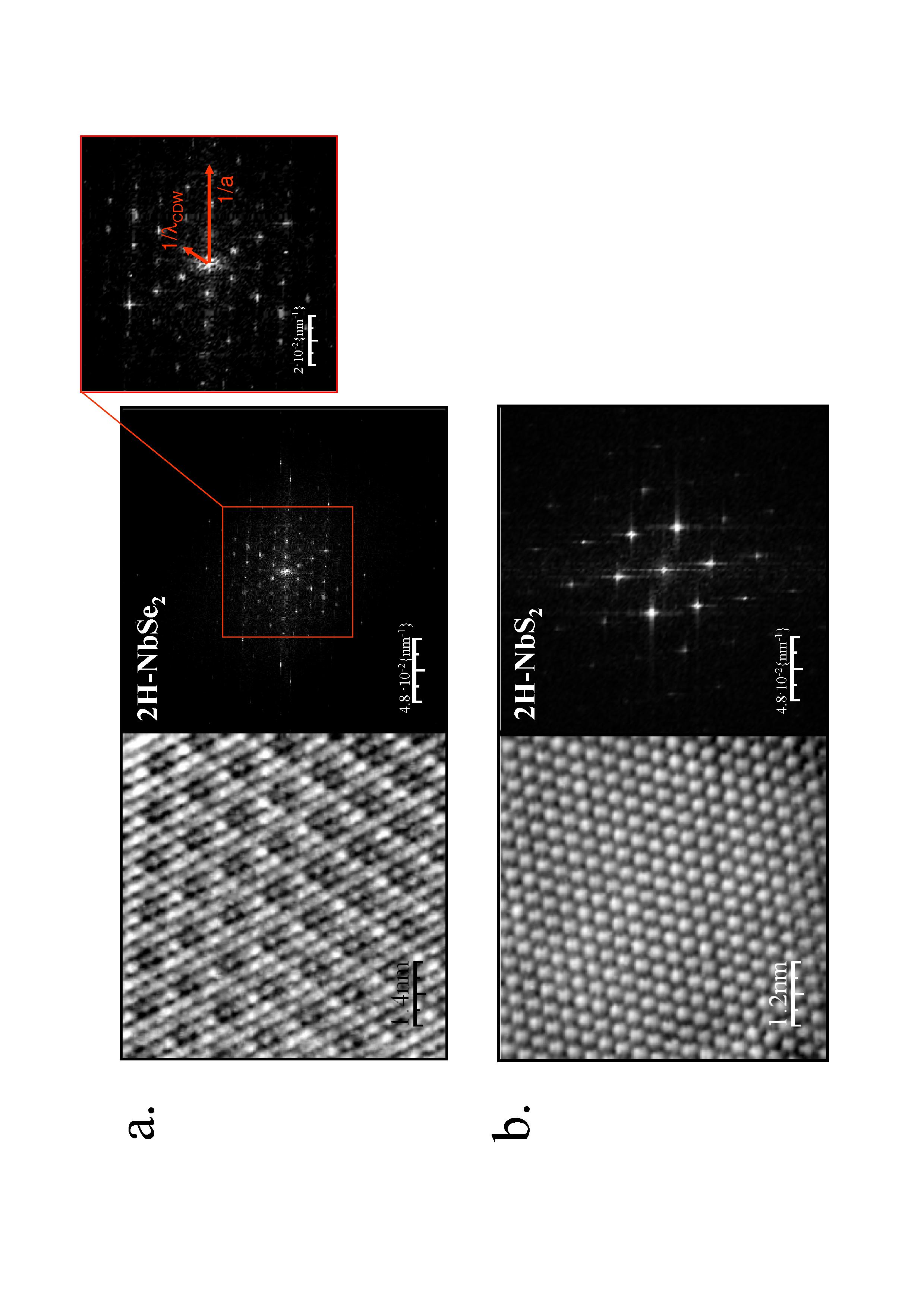}
\caption{{\small{a) Atomic scale topography (left panel), taken with STM in 2H-NbSe$_{2}$ at 0.1K. A superimposed modulation due to the presence of the CDW is observed. The corresponding Fourier transform image (right panel) shows the wave vectors of the atomic and CDW modulations in the reciprocal space (marked by arrows in the inset). b) Atomic scale topography at 0.1 K (left panel) and its Fourier transform image (right panel) at atomic scales in $2H-NbS_{2}$. No CDW order is observed.}}\label{fig1}}
\end{figure}

STM images in 2H-NbS$_2$ show atomically flat regions with lateral sizes that easily go up to several hundreds of nanometers. In fig.\ \ref{fig1} we present a STM image with atomic resolution obtained at 0.1 K in 2H-NbS$_{2}$ (fig.\ \ref{fig1}b), and compare it to a STM image in 2H-NbSe$_{2}$ measured under the same conditions (fig.\ \ref{fig1}a). In 2H-NbSe$_{2}$, we observe the hexagonal surface atomic lattice of Se atoms and the CDW order as previously reported in many other STM experiments and largely discussed in literature\ \cite{Sacks98,coleman88}. In 2H-NbS$_{2}$, the STM image (fig.\ \ref{fig1}b) also shows the hexagonal symmetry of the surface atomic lattice of the chalcogenide (here S), characteristic of this family of materials, with a slightly smaller lattice constant, as expected from X-Ray measurements\ \cite{Fisher80,Hamaue86,Meerschaut01}. However, both the STM image and its associated Fourier transform in 2H-NbS$_2$ (left and right panels in fig.\ \ref{fig1}b, respectively) do not show any modulations besides of the ones coming from atomic periodicity. This result gives strong evidence that there is no coexistence of superconducting and CDW orders in 2H-NbS$_2$ at least down to 0.1 K.

\begin{figure}[ht]
\includegraphics[angle=0,width=8.5cm,clip]{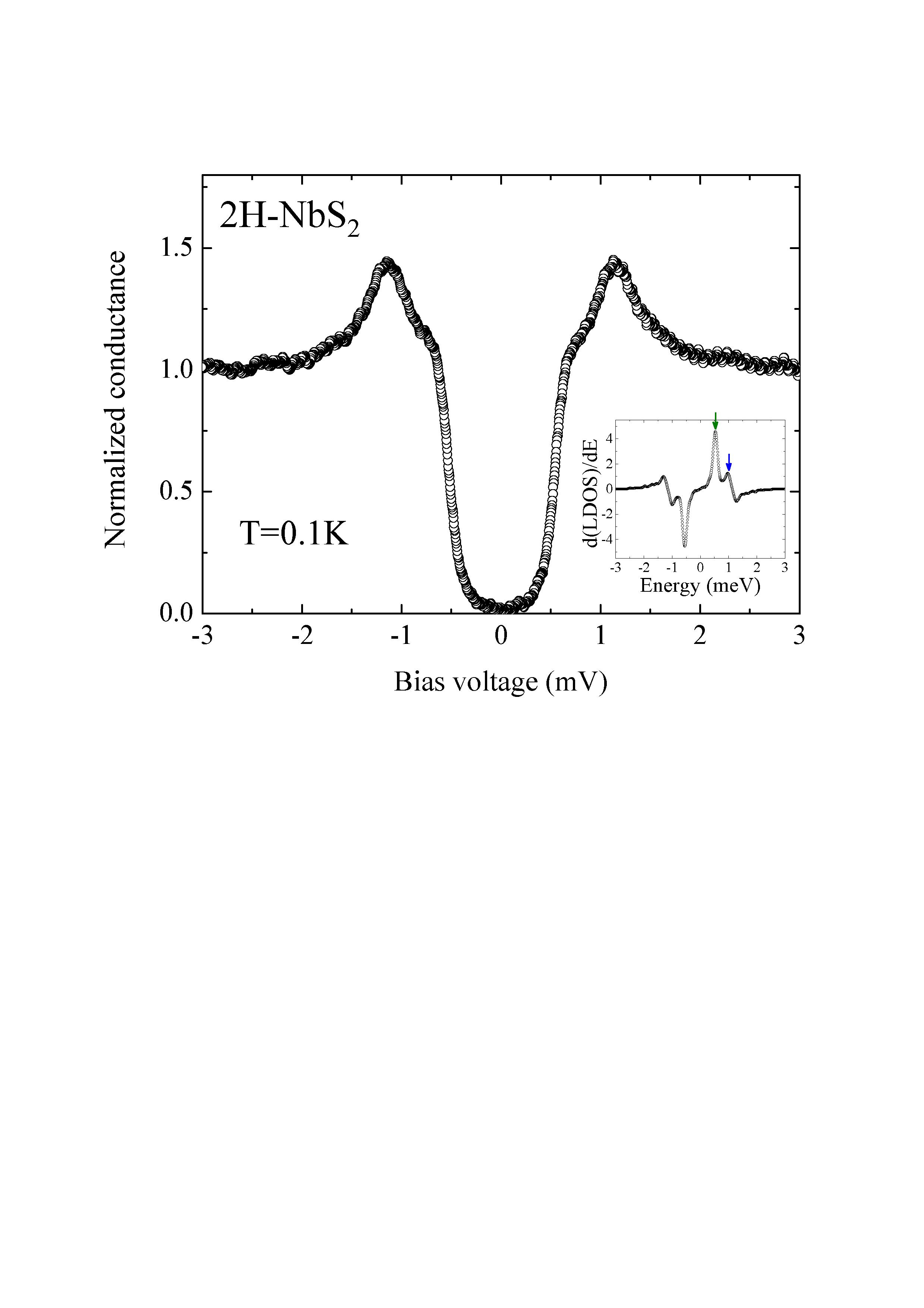}
\vskip -3 cm
\caption{{\small{Characteristic tunneling conductance curve obtained in 2H-NbS$_{2}$ and measured at 0.1K. The derivative of the associated LDOS (shown in the inset) presents two peaks, at $0.97 meV$ and $0.53 meV$ (marked by blue and green arrows, respectively), which correspond to two peaks in a gap distribution.}}\label{fig2}}
\end{figure}

\begin{figure}[ht]
\includegraphics[width=8cm,clip]{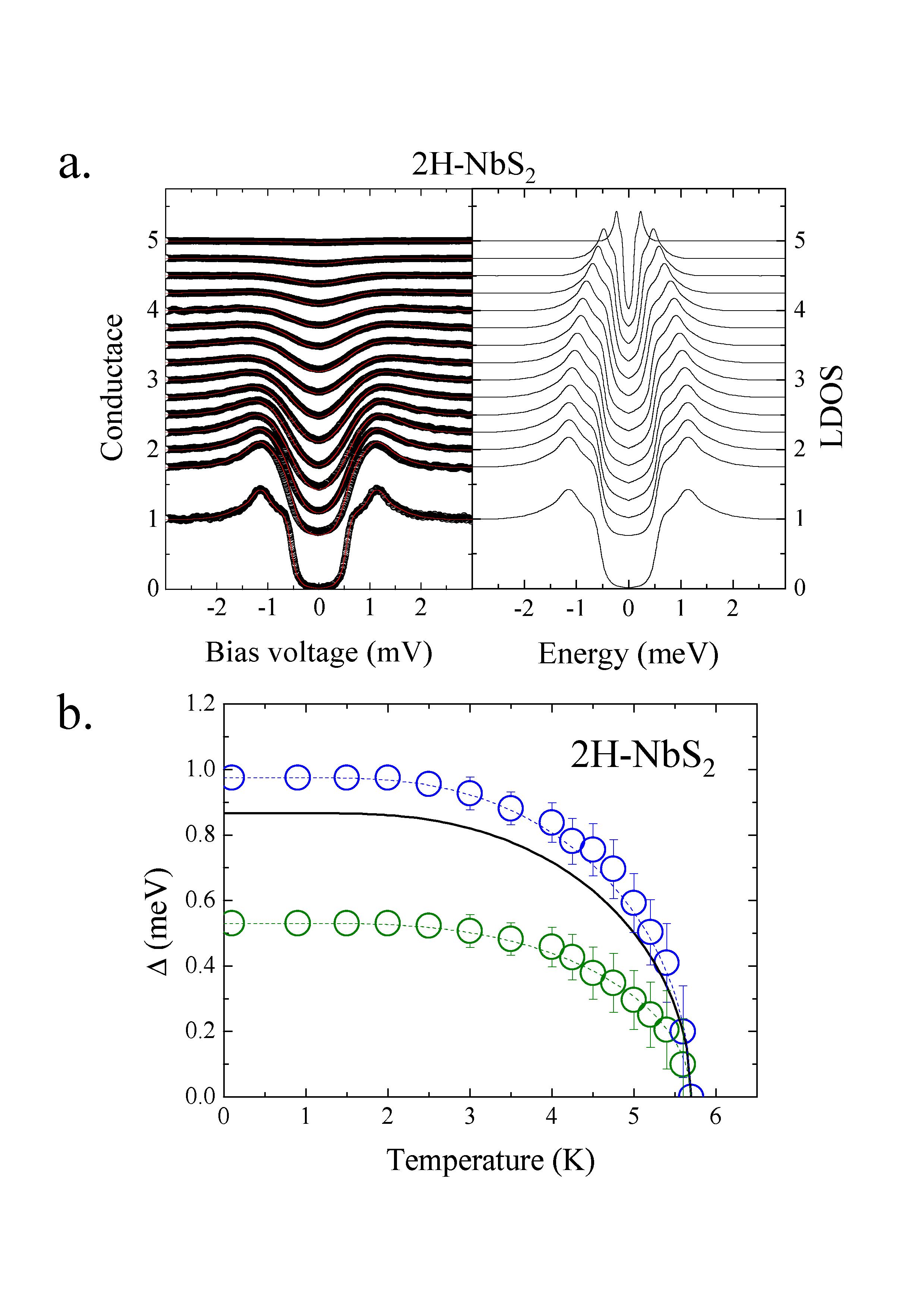}
\caption{{\small{a) Temperature scan of tunneling conductance spectra (left) and superconducting LDOS (right). The LDOS is obtained by deconvolution from tunneling conductance curves.  The data are, from bottom to top, taken at 0.1K, 0.9K, 1.5K, 2K, 2.5K, 3K, 3.5K, 4K, 4.25K, 4.5K, 4.75K, 5K, 5.2K, 5.4K and 5.6K. Red lines in left panel are the fits obtained from convoluting the LDOS shown in the right panel with temperature. b) Temperature dependence of the two peaks in $d(LDOS)/dE$ (green and blue points, corresponding to the features marked by green and blue arrows in fig.\ \protect\ref{fig2}, dashed lines are guides to the eye). The solid line is the BCS expression taking $\Delta = 1.76 k_{B}T_{c} = 0.87$ meV with $T_{c}$ = 5.7 K.}}\label{fig3}}
\end{figure}

In fig.\ \ref{fig2} we show a typical tunneling conductance vs. bias voltage curve obtained in an atomically flat region at 0.1 K. The curve is fully reproducible over the whole surface, showing tiny differences (below 10 \%) as a function of the position at atomic scale, in a similar way as found in 2H-NbSe$_2$ previously \ \cite{Guillamon08}. As is well known, the local tunneling conductance simply reflects the convolution between the LDOS of the sample and the derivative of the Fermi function. At these temperatures, the tunneling conductance curve pretty directly follows the LDOS, because the derivative of the Fermi function is very narrowly peaked, in particular as compared to the critical temperature (here $0.1K \approx 0.02 T_{c}$). The LDOS is very far from the divergent quasiparticle peaks expected within single band s-wave BCS model, and shows instead a slight increase from relatively low energies, of about 0.3 meV, which becomes much more marked around 0.6 meV, and a rounded quasiparticle peak somewhat above 1 meV. Such a LDOS evidences that more than a single gap is open over the Fermi surface, i.e. there is distribution of values of the superconducting gap. It is easily shown that $d(LDOS)/dE$ has peaks at energies which correspond to the superconducting gaps which appear more often in the distribution. As shown in the inset of fig.\ \ref{fig2}, two well defined peaks $d(LDOS)/dE$ are visible at energies $\pm (0.97 \pm 0.03)$ meV and $\pm (0.53 \pm 0.03)$ meV. The temperature dependence of the tunneling conductance and of the LDOS (obtained as in Ref.\ \cite{Crespo06a}) is shown in fig.\ \ref{fig3}a. The LDOS has a smooth temperature dependence. The features appearing at low temperatures are simply scaled to lower energies as $T_c$ is approached. The temperature dependence of the two peaks in $d(LDOS)/dE$ is shown in fig.\ \ref{fig3}b. Both of them disappear at $T_{c}$=5.7\ K. Moreover, the value of the gap expected within single gap s-wave BCS model (1.76 k$_BT_{c}$=0.87 meV; the solid line in fig.\ \ref{fig2}b gives its temperature dependence), is in between the values of the two peaks on $d(LDOS)/dE$ at all temperatures (fig.\ \ref{fig2}b). This gives strong evidence that two different superconducting gaps ($\pm (0.97 \pm 0.03)$ meV and $\pm (0.53 \pm 0.03)$ meV) open at different parts of the Fermi surface of 2H-NbS$_2$.

The situation resembles the results found previously in MgB$_{2}$ \ \cite{Rubio01,Schmidt02,Szabo01,Giubileo01,Liu01,Eskildsen02}, and in 2H-NbSe$_2$ \ \cite{Boaknin03,Rodrigo04c,Suderow05d,Fletcher07,Guillamon08,Yokoya01,Kiss07}, which evidenced the appearance of two band superconductivity in these materials. In 2H-NbS$_2$ the gap values appear to be much closer to each other than in the case of MgB$_2$, where the two gaps differ by more than a factor of three. Moreover, interband scattering effects should be present to allow for a smooth temperature dependence of both gaps\ \cite{Suhl59}. However, in 2H-NbS$_2$ the two gap nature seems better developed than in 2H-NbSe$_2$, where similar features in the density of states are less clearly resolved \ \cite{Hess90,Rodrigo04c,Suderow05d}.

\begin{figure}[ht]
\includegraphics[width=7.5cm,clip]{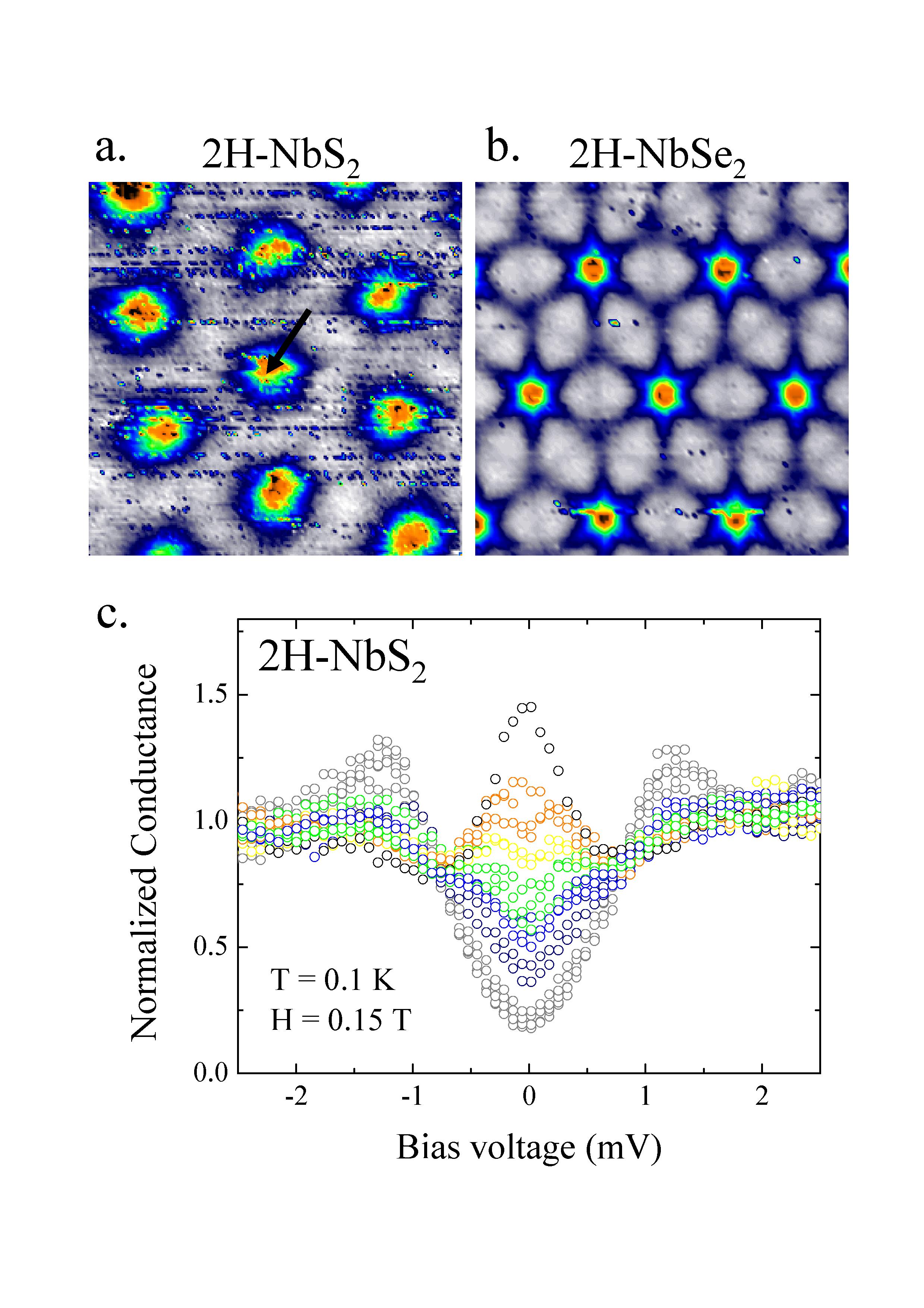}
\caption{{\small{a) STS image of the vortex lattice in 2H-NbS$_{2}$ taken at 0.1 K and 0.15 T is shown in a (360 nm $\times$ 360 nm). In b, we show, for comparison, a STS image of the same area obtained in 2H-NbSe$_2$ under similar conditions. In c we present tunneling conductance curves in 2H-NbS$_2$ along a line of about 65 nm that goes from the middle point between vortices into the vortex core (using a similar color scheme as in a and b), following the line shown in a.}}\label{fig4}}
\end{figure}

Under magnetic field we observe an ordered hexagonal vortex lattice in atomically flat regions for different values of the applied field. A STS image of the vortex lattice at 0.1 K and 0.15 T built from the conductance curves at zero bias voltage is shown in the fig.\ \ref{fig4}a, and compared to results at the same field and temperature in 2H-NbSe$_2$ (fig.\ \ref{fig4}b)\ \cite{Hess89,Hess90,Hess91}. The vortex core shapes, as observed with STS, are dramatically different in both compounds. First, the star shape of 2H-NbSe$_2$ is fully absent in 2H-NbS$_2$. When making STS images at different energies, vortex cores in 2H-NbS$_2$ appear to be round, fully respecting angular symmetry, at all energies, whereas 2H-NbSe$_2$ has a rich phenomenology, which includes the formation of complex angular patterns, discussed in detail in previous works\ \cite{Hess89,Hess90,Hess91}. On the other hand, the energy dependence of the LDOS at the center of the vortex core in 2H-NbS$_2$ is very similar to the one found in 2H-NbSe$_2$. There is a high peak at the Fermi level, due to the lowest level of bound localized states inside the vortex core wells\ \cite{Caroli64}, which splits when going out to the vortex core, exactly as in 2H-NbSe$_2$. The difference between both compounds lies only in the angular dependence of these features in the LDOS. Whereas in 2H-NbSe$_2$, the way the localized state peak splits is complex and shows a very strong real space angular dependence, in 2H-NbS$_2$, the LDOS respects in-plane symmetry. This must be correlated to the absence of CDW in 2H-NbS$_2$ (Fig.1), so we can state that the peculiar star shape of the Abrikosov vortex lattice in 2H-NbSe$_2$ and its voltage dependence is due to the presence of CDW order in this compound.

Note that two band superconductivity appears to be common to 2H-NbSe$_2$ and to 2H-NbS$_2$, and it could be a rather general feature of this family of materials which deserves further study. For example, two different gaps may open at the two groups of transition metal d-electron derived FS sheets\cite{Johannes06}. However, 2H-NbSe$_2$ has an additional gap modulation caused by CDW that shapes the angular dependent properties of the different superconducting gaps opened over one of the bands, or eventually over the whole Fermi surface. Indeed, different models directly associate the star shape in the vortex core localized state features in 2H-NbSe$_2$  with a strong in-plane modulation of the superconducting gap\ \cite{Gygi90,Hayashi96,Nakai06}. The superconducting properties of 2H-TaS$_2$ and of 2H-TaSe$_2$, where CDW order appears at a much higher temperature (80 K and 120 K respectively), could be even more intriguing and may also include a richly shaped vortex lattice.

On the other hand, naturally, the presence of features in the LDOS due to localized bound states strongly depends on the mean free path, as scattering flattens out the peak in the LDOS at the center of the vortex core\ \cite{Renner91,Bergeal06}. So the observation of the peak in the LDOS in 2H-NbS$_2$ at the center of the cores shows that our samples are within or close to the clean limit. Also, 2H-NbS$_2$ appears as the system that best approaches the initial proposal of Caroli, de Gennes and Matricon regarding quasiparticle bound states inside cores\ \cite{Caroli64}. Indeed, whereas 2H-NbSe$_2$ is strongly influenced by CDW, vortex core features in other systems, as e.g. high $T_c$ cuprate and related superconductors, boron doped diamond, or YNi$_2$B$_2$C have a rather rich and intriguing behavior \ \cite{Fischer07,Sacepe06,Nishimori04}, with, however, no clear bias symmetric peak appearing at the Fermi level LDOS at the center of the cores.

In summary, we have found that the LDOS of 2H-NbS$_2$ has two pronounced features related to intrinsic two gap superconductivity, in close analogy to MgB$_2$. Vortex cores show a bound state close to the Fermi level and do not present any in-plane anisotropy. By comparing 2H-NbS$_{2}$ with previous work on 2H-NbSe$_{2}$, we find that charge order is the main responsible of the strong in-plane six-fold anisotropy that occurs in 2H-NbSe$_{2}$. In particular, CDW is responsible for the six fold structure of the vortex lattice found by Hess et al.\ \cite{Hess90} in 2H-NbSe$_2$.

\section{Acknowledgments.}
We acknowledge J. Martial at IMN for her help in sample preparation, and discussions with A. Mel'nikov, A.I. Buzdin, F. Guinea, J.G. Rodrigo, V. Crespo and J.P. Brison. The Laboratorio de Bajas Temperaturas is associated to the ICMM of the CSIC. This work was supported by the Spanish MEC (Consolider Ingenio 2010, MAT and FIS programs), by the Comunidad de Madrid through program "Science and Technology in the Millikelvin", by Grant No. ANR-ICENET NT05-1 44475, and by NES and ECOM programs of the ESF.

\bibliographystyle{apsrev}

\end{document}